\documentclass[11pt]{article}
\usepackage[utf8]{inputenc}
\usepackage{graphicx}
\usepackage{url}
\usepackage{fullpage}
\usepackage{hyperref}
\hypersetup{colorlinks,linkcolor=black,filecolor=black,urlcolor=black,citecolor=black}
\usepackage[small,bf]{caption}
\title{A note on sampling biases in the Bangladesh mask trial}
\author{Maria Chikina{$^\ast$}, Wesley Pegden{$^\dagger$}, Benjamin Recht{$^\#$}\\
$\ast$ University of Pittsburgh\\
$\dagger$ Carnegie Mellon University\\
$\#$ University of California, Berkeley
}
\date{December 2, 2021}

\begin{document}

\maketitle

\begin{abstract}
    A recent cluster trial in Bangladesh randomized 600 villages into 300 treatment/control pairs, to evaluate the impact of an intervention to increase mask-wearing.  Data was analyzed in a generalized linear model and significance asserted with parametric tests for the rate of the primary outcome (symptomatic and seropositive for COVID-19) between treatment and control villages.
    
    In this short note we re-analyze the data from this trial using standard non-parametric paired statistics tests on treatment/village pairs.  With this approach, we find that behavioral outcomes like physical distancing are highly significant, while the primary outcome of the study is not.  Importantly, we find that the behavior of unblinded staff when enrolling study participants is one of the most highly significant differences between treatment and control groups, contributing to a significant imbalance in denominators between treatment and control groups.
    
    The potential bias leading to this imbalance suggests caution is warranted when evaluating rates rather than counts.  More broadly, the significant impacts on staff and participant behavior urge caution in interpreting small differences in the study outcomes that depended on survey response.
\end{abstract}
\section{Introduction}

A recent cluster RCT of mask promotion in Bangladesh \cite{bang} reported decreases in symptomatic seroprevalence (primary outcome), decreases COVID-19-like symptoms without antibody confirmation (secondary outcome), and increases in mask wearing behavior (secondary outcome).  The preprint describing the results of this trial analyzes the data using a generalized linear model and finds roughly a 10\% decrease in the fraction of the enrolled population which ends up with a positive test for COVID-serology after reporting COVID-like symptoms. The paper’s approach evaluates this result as significant at p=0.05.

\begin{figure}[t]
    \centering
    \includegraphics[width=\linewidth]{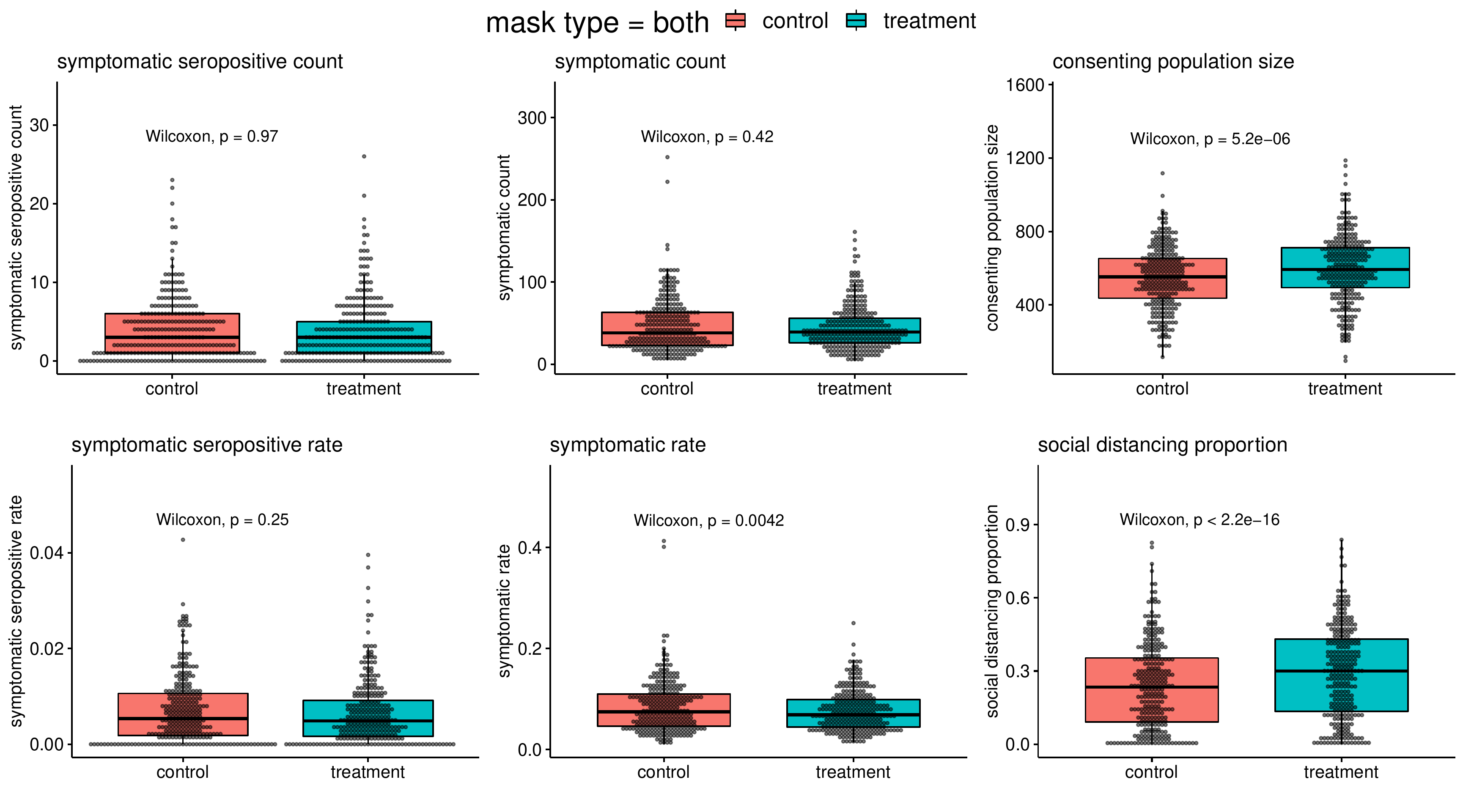}\\
    \caption{Differences in treatment vs control groups.  Significance is evaluated with Wilcoxon paired tests.}
    \label{fig:both}
\end{figure}

In this note we re-analyze the data released from this trial using non-parametric paired tests, and illustrate some basic properties of the data, highlighting possible alternative explanations for the differences between the treatment and control groups. In particular, we note a lack of blinding of the observers in the study contributed to highly significant differences between the population of the treatment and control groups. These differences are much more significant than the measured differences in symptomatic seropositivity.

\section{Study Protocol}
\label{sec:protocol}
The cluster RCT was run at the village level. 600 villages in Bangladesh were paired based on COVID-case data, population density, and population size. Each paired village was assigned to treatment or control at random. As described in the the paper accompanying the study, households were enrolled in the study in a 2-step process.  In the first step, blinded staff members mapped villages and household locations.  In the second step, unblinded staff members sought consent from eligible household members; unblinded staff recorded for each household that they (a) consented to participate, (b) declined to participate, or (c) were ``unreachable''.  The study then proceeded to implement a mask promotion intervention in the treatment villages. In both villages, participants were asked to report COVID-like symptoms. Those who reported symptoms were asked to volunteer blood draws for serology. The primary endpoint was evaluated based on the number of these blood draws that tested positive for COVID antibodies.

Among 300 matched village pairs, there were 200 pairs in which the treatment village was assigned to surgical masks, and 100 pairs in which the treatment group was assigned to cloth masks.

\section{Impact of trial intervention}
Despite 300:300 randomization of the 600 villages, there was a notable imbalance in the size of the consenting populations between the control and treatment groups, which we explore below. The control group contained 156,938 individuals while the treatment group contained 170,497 individuals. The total absolute numbers of symptomatic seropositives in the treatment and control villages was 1086 and 1106, respectively. This difference is too small to be significant even if participants had been individually randomized ($p=0.34$ in a Binomial test for probability $\frac 1 2$); in the study, an effect is asserted for the \emph{rate} of symptomatic seropositives, i.e., normalized by population denominators. We note that the 10\% decrease reported in the fraction of individuals who become symptomatic seropositives is not driven primarily by a ($\approx 2\%$) decrease in the numerator of this fraction of symptomatic seropositives, but instead an ($\approx 9\%$) increase in the denominator.

Why was the denominator different? Other than the impact on mask-wearing itself and increases in physical distancing, the greatest difference between the control and treatment arms was on the rate of households approached to be included in the trial data, and thus included the denominator of these rates.  In paired Wilcoxon tests, %(https://en.wikipedia.org/wiki/Wilcoxon_signed-rank_test)
the impact on the overall rate of approaching households is significant at $p=10^{-11}$, while neither the count nor the rate of symptomatic seropositives (the primary study outcome) is significant using the same analysis.  The rate, but not the count, of reported COVID-like symptoms (unconfirmed by antibody testing) is significant at p=0.004 (Figure \ref{fig:both}). Overall, we see large intervention effects on aspects of behavior (i.e., mask-wearing and physical distancing), but much smaller effects on symptomatic seropositivity, which do not pass significance thresholds using the same nonparametric tests against which we find that behavior and enrollment differences are highly significant between treatment and control groups.

\begin{figure}[t]
    \centering
    \includegraphics[width=\linewidth]{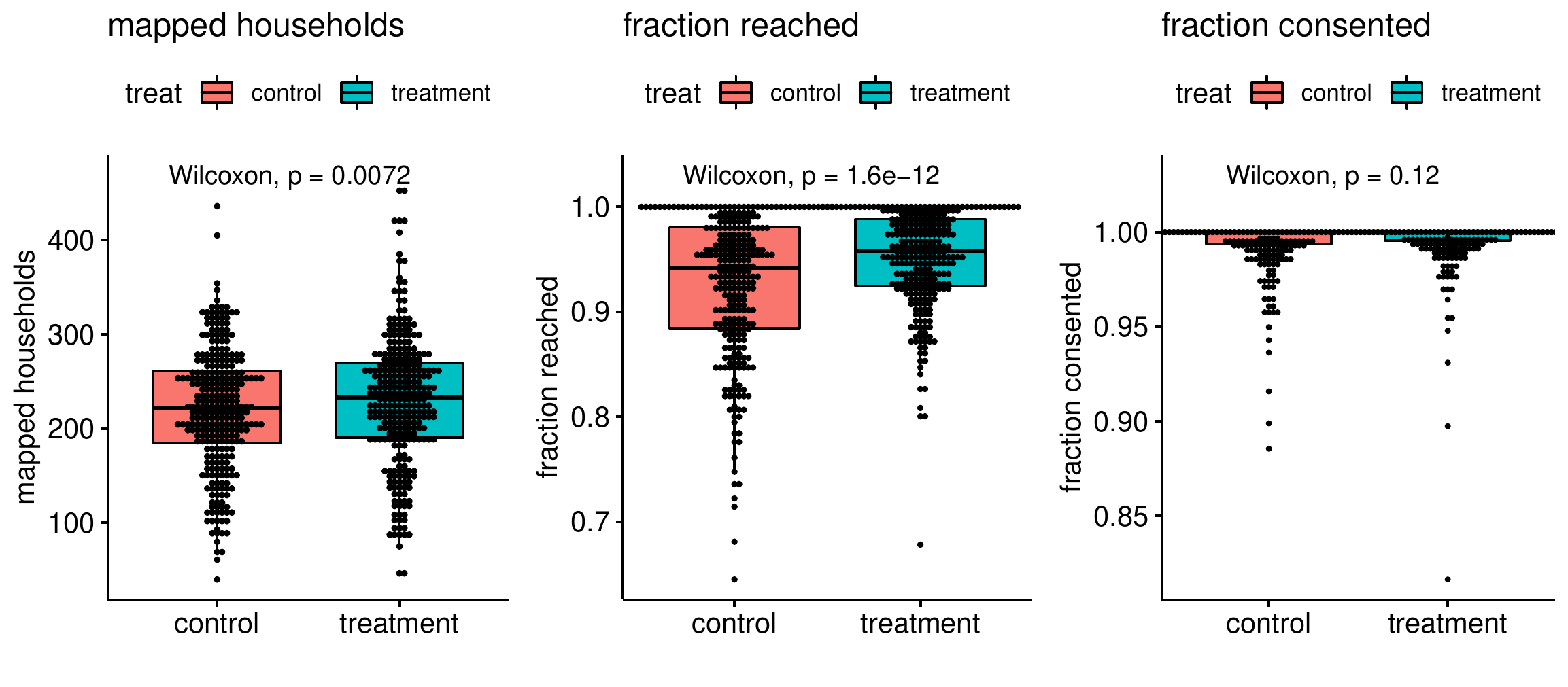}\\
    \caption{Among all study variables, one of the largest differences between treatment and control groups arises in the step when unblinded staff sought consent from households to be included in trial data.  The number of mapped households (left) is the number of households in each village mapped by blinded staff in the first step of enrollment.  The fraction of households reached is the fraction of mapped households that either agreed or declined to consent, rather than being deemed unreachable by the staff member due to ``not available, absent, temporarily or permanently migrated, or other.'' The fraction of consenting households represents the fraction of those who were reached that agreed rather than declined to consent. Significance is evaluated with Wilcoxon paired tests.}
    \label{fig:consent}
\end{figure}

\begin{figure}[p]
    \centering
    \includegraphics[width=\linewidth]{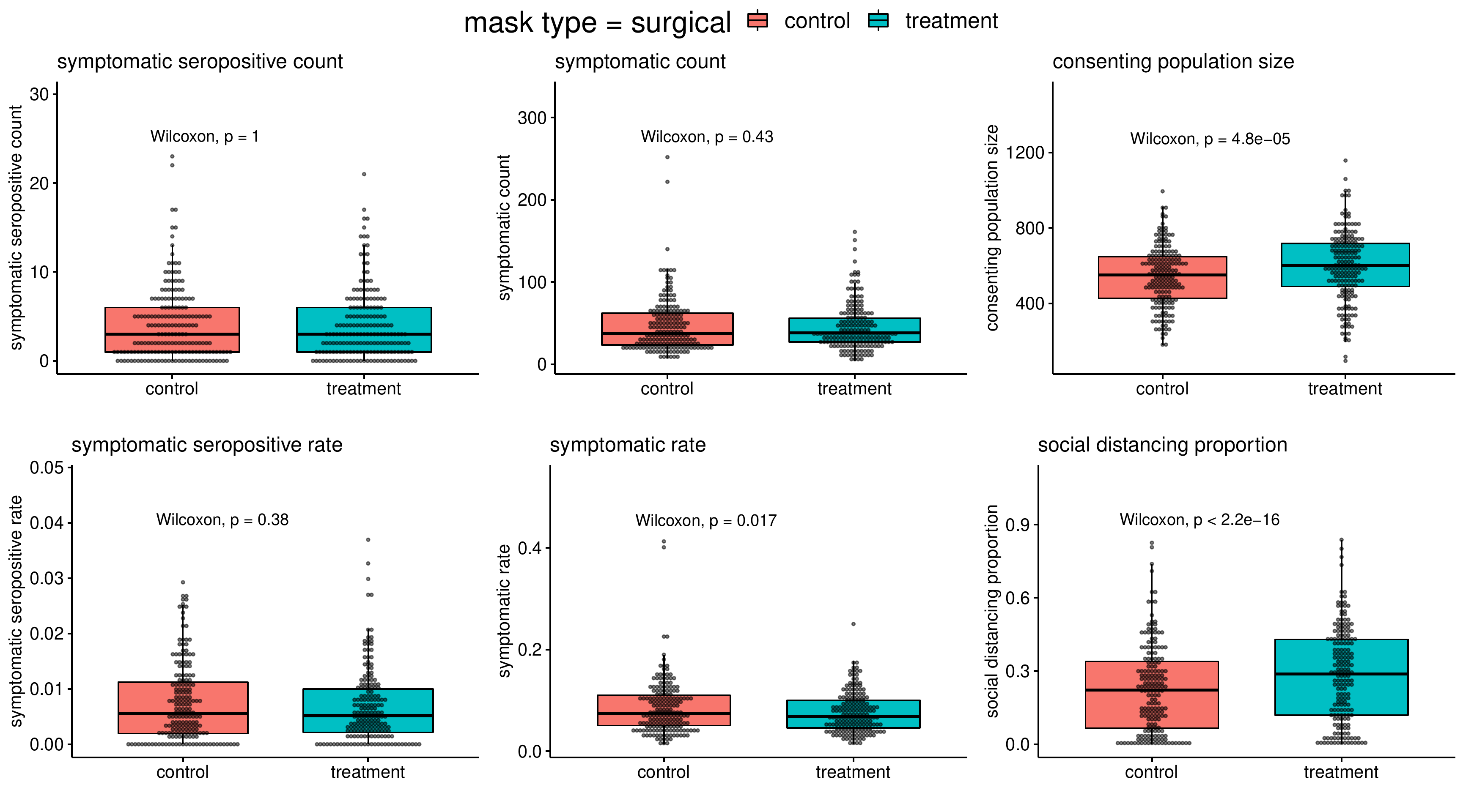}\\
    \caption{Differences in treatment vs control groups, restricted to surgical village pairs. Significance is evaluated with Wilcoxon paired tests.}
    \label{fig:surgical}
\end{figure}

\begin{figure}[p]
    \centering
    \includegraphics[width=\linewidth]{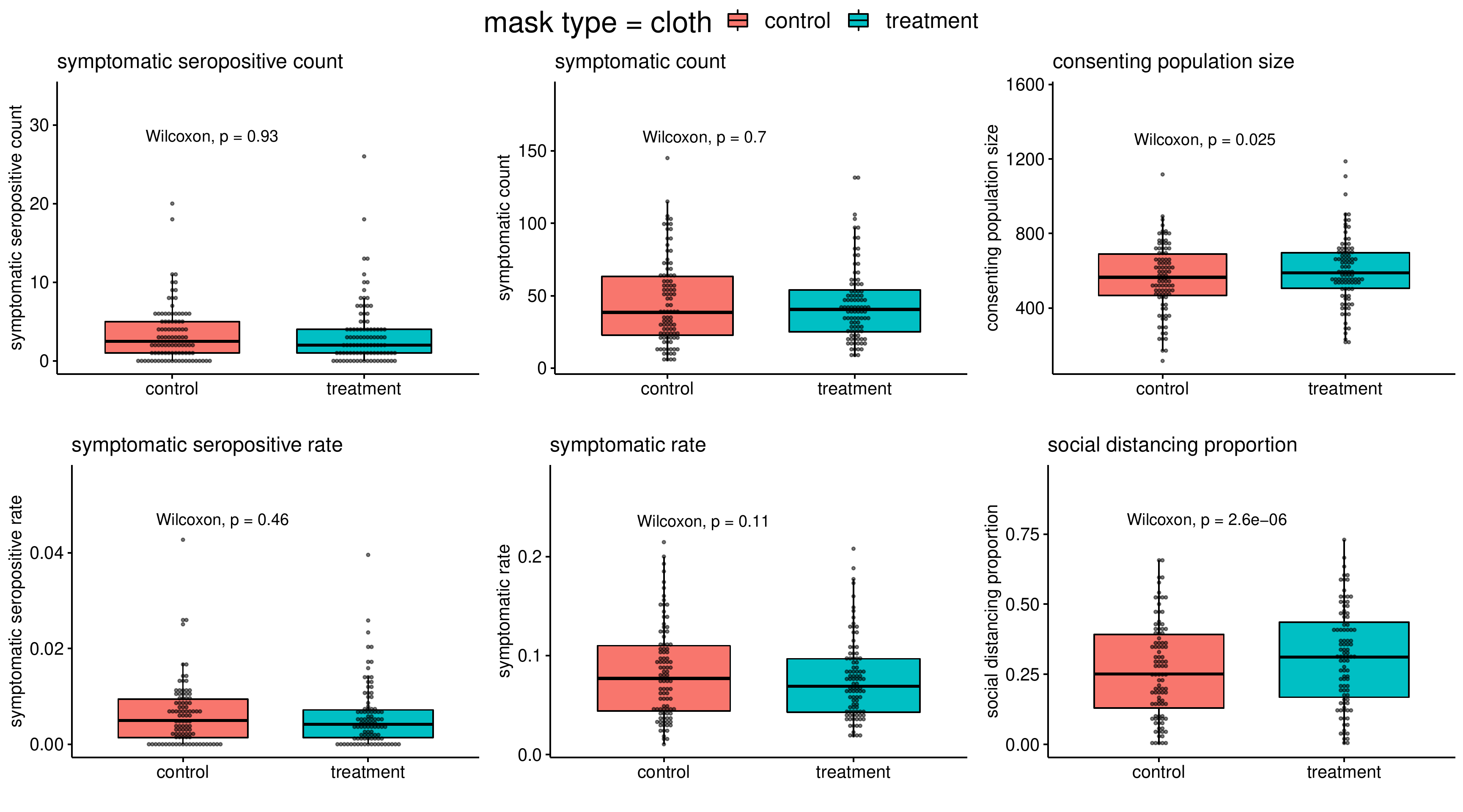}\\
    \caption{Differences in treatment vs control groups, restricted to cloth village pairs. Significance is evaluated with Wilcoxon paired tests.}
    \label{fig:cloth}
\end{figure}

\section{Contributors to population imbalance}

As described in Section \ref{sec:protocol}, enrollment took place in a 2-step process, with blinded staff mapping households, and unblinded staff seeking consent from those households.  Roughly $4.5\%$ more households were mapped in treatment villages than control villages in the first step.  This difference in mapping behavior between treatment and control groups is mildly significant at $p=0.0072$ (left panel, Figure \ref{fig:consent}), despite supposed staff blinding\footnote{It is notable that the randomization procedure for the trial did not uniformly randomly choose a final treatment/control designation among matched village pairs.  Instead, it re-randomized 50 times, choosing a “best” split with respect to balance for case trajectories and case rates; see page 22 of the supplementary appendix in \cite{bang}.  While blinding issues are perhaps the most natural factor to suspect if the $p=0.007$ split in mapped households is not to be seen as a rare coincidence, it is interesting to speculate whether the conditioning on a 2\% event in the sampling procedure---even one which is intended to improve the balance of groups with respect to certain traits---could increase the likelihood that some apparently significant events of interest happen by chance.  Note that in the worst case, this could demand tightening significance thresholds by at most a factor of 50, which means that significance we see for the impact of the intervention on behaviors like mask-wearing, distancing, and rates of consent would survive.}.
In Figure \ref{fig:consent}, we see that the behavior of unblinded staff members in the second step of the process is highly significant at $p=10^{-11}$.  This further increases the final imbalance in the number of households between treatment and control groups to $8\%$, with total populations differing between treatment and control groups by $9\%$, and is the single most significant effect of the intervention we see on variables other than rates of mask-wearing and physical distancing.  Among approached households, consent rates are not significantly different (right panel).

\section{Potential impact of imbalance}

The strong effect of the intervention on the behavior leading to baseline consent is especially notable, since inferring causal effects in the presence of a strong effect on enrollment rates requires assuming that  ``borderline participants'' who would be have been consented to participate in the treatment but not control groups were just as likely as typical villagers to become infected with COVID, develop symptoms, and report them to study staff to become eligible for endline blood draws. Adopting a conservative intention-to-treat framework \cite{intent} (“once randomized, always analyzed”), one could look for significant effects on the overall counts of symptomatic seropositivity, rather than leveraging the larger consenting denominator in treatment groups to estimate the effect of treatment on rates of symptomatic seropositivity.  But these counts are essentially balanced in the raw data of the trial (1086 vs 1106 symptomatic seropositives in treatment vs control) and in subgroups we examined.  The authors of \cite{bang} carry out analyses to check robustness of their results to how they deal with missing data from participants who report symptoms but decline to participate in endline blood draws.  The same robustness is not checked for the extreme imbalance we observe in the patterns of who was initially approached to participate in the study.

\section{Conclusion}

Overall, this simple analysis suggests that the impact of the mask intervention was highly effective at modifying behaviors (distancing, mask-wearing, symptom reporting), but that any effect on actual symptomatic seropositivity was much more subtle; in particular, whatever effects the intervention had on the rate of symptomatic seropositivity in the villages was certainly not large relative to other factors contributing to variance in this parameter across villages.  For example, it might seem surprising, even if one did not expect any effect from masks at all, that intervention's impact on behaviors alone (which extended to highly significant effects on physical distancing) would not translate to clearer differences in symptomatic seropositivity between treatment and control groups.

We suggest that the very large causal effects on consent rates and thus population denominators urge caution in interpreting the small differences we see in symptomatic seropositivity between treatment and controls, which are already not statistically significant according to standard nonparametric paired tests.  Additionally, as the trial shows that the intervention studied can have large and highly significant effects even on unintended aspects of behavior, including staff surveying behavior, bias-susceptible endpoints that depend on subjective reports of symptoms from participants to staff should be used with care.  

\bigskip

\noindent Code to reproduce our figures is available at \url{https://github.com/mchikina/maskRCTnote}.

\section{Appendix: Notes on subgroup analyses}

\begin{figure}[p]
    \centering
    \includegraphics[width=\linewidth]{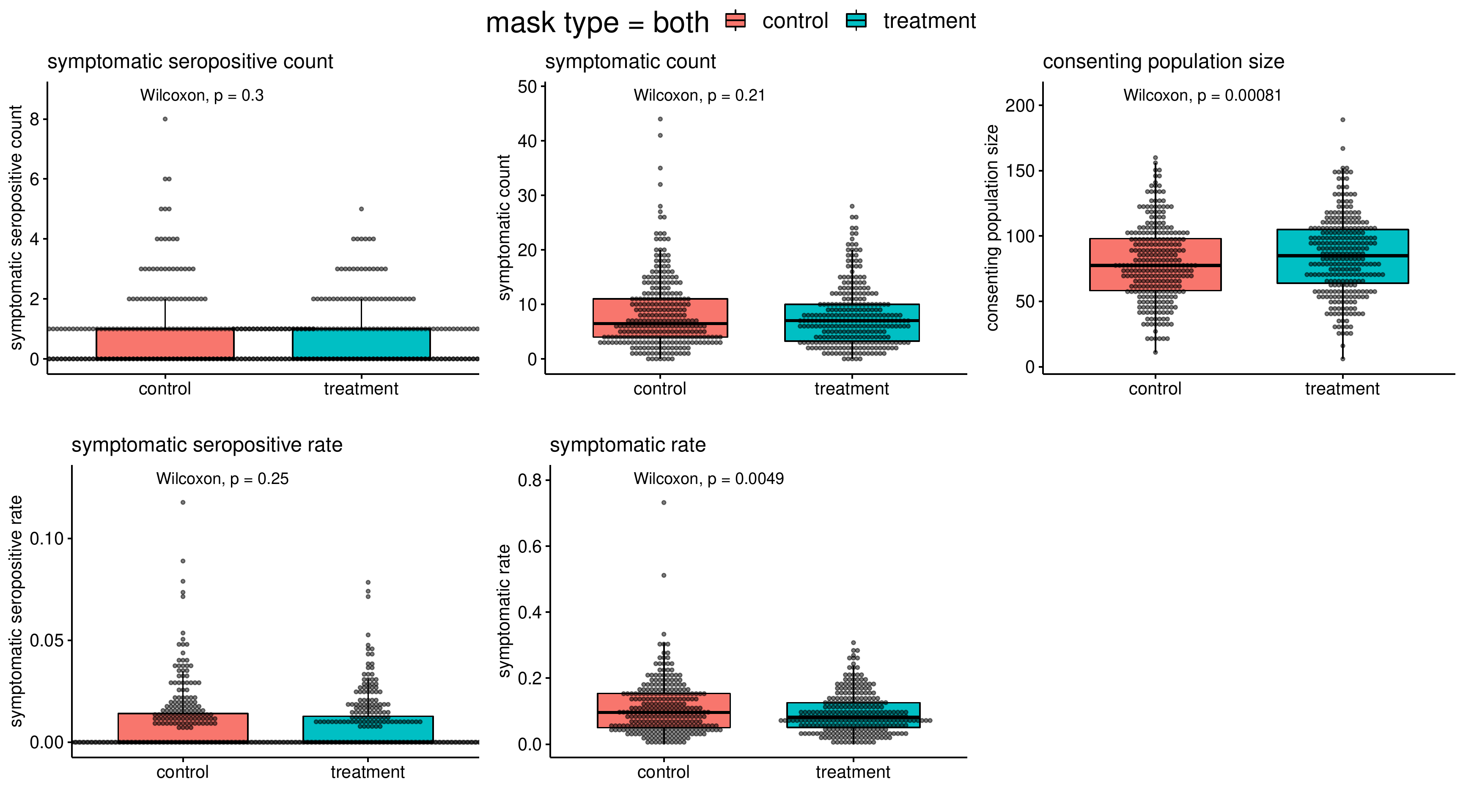}
    \caption{Difference between treatment and control groups when restricted to over-60 village populations.  Significance is evaluated with Wilcoxon paired tests.}
    \label{fig:over60}
\end{figure}

\begin{figure}[p]
    \centering
    \includegraphics[width=\linewidth]{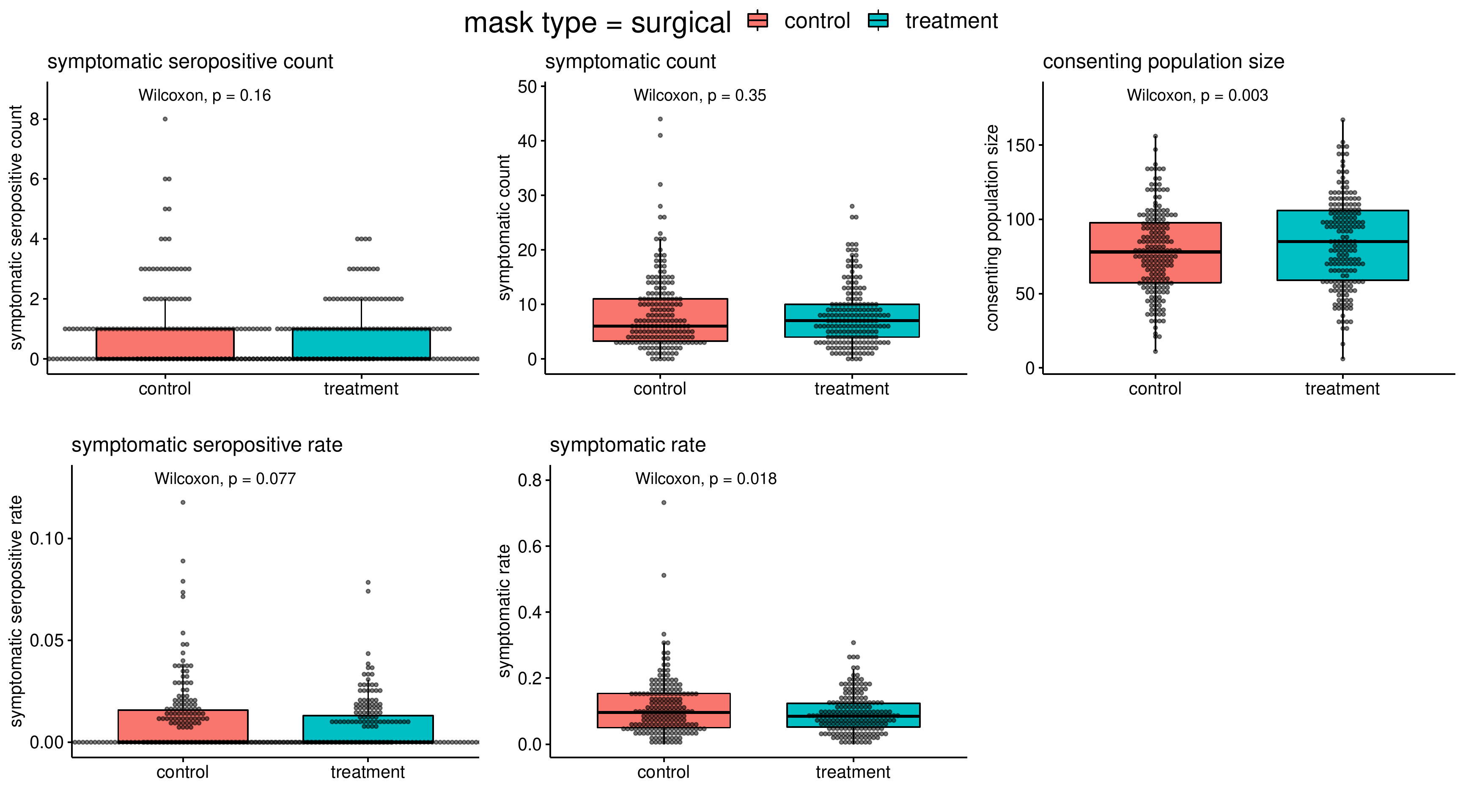}
    \caption{Difference between treatment and control groups when restricted to over-60 village populations in villages assigned to surgical masks. Significance is evaluated with Wilcoxon paired tests.}
    \label{fig:over60-surgical}
\end{figure}

\begin{figure}[p]
    \centering
    \includegraphics[width=\linewidth]{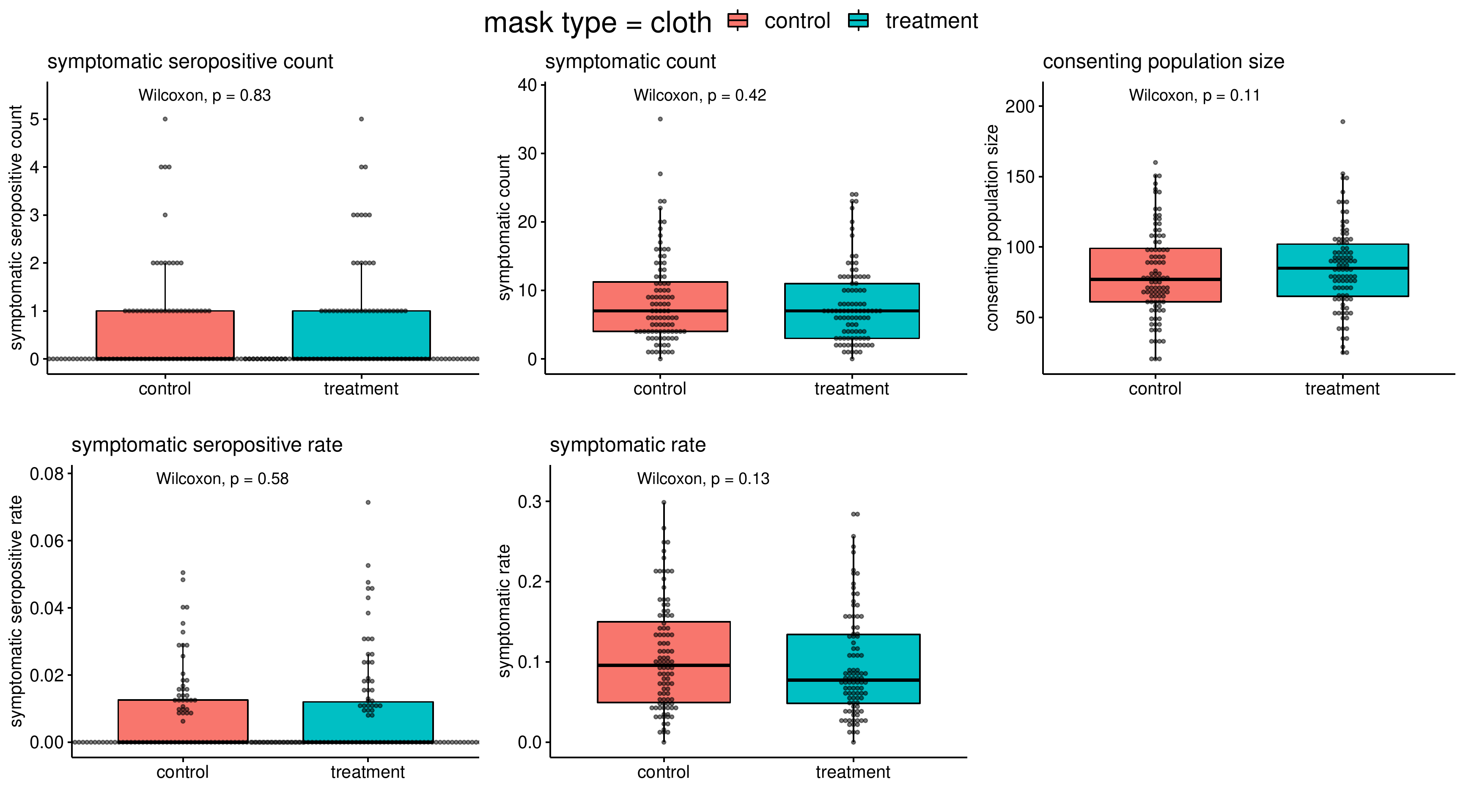}
    \caption{Difference between treatment and control groups when restricted to over-60 village populations in villages assigned to cloth masks. Significance is evaluated with Wilcoxon paired tests.}
    \label{fig:over60-cloth}
\end{figure}

In \cite{bang}, the greatest effect on and significance for symptomatic seropositivity is claimed for surgical masks in older adults.  In Figure \ref{fig:over60}, we show data restricted to the over-60 population of each village.  Figures \ref{fig:over60-surgical} and \ref{fig:over60-cloth} show panels isolating villages assigned to surgical and cloth masks respectively. In every case, behavioral and denominator-size differences are the most significantly different between treatment and controls.   Among all village pairs and surgical village pairs, the rate of reported (not test confirmed) symptoms is significant, while the count is not. For the primary outcome of sympomatic seropositivity, the smallest $p$-value we see for the rate is $p=0.077$, and for the count is $p=0.16$.   Note that the study data allows retrospective subset analysis for several age groups (18-30, 30-40, 40-50, 50-60, and over 60's, among others), so significance thresholds for subset analyses must be treated with some caution.

\end{document}